\documentclass[aps,pre,preprint,onecolumn,showpacs,amsmath,amssymb]{revtex4-1}
\usepackage[T1]{fontenc}
\usepackage{amsmath}
\usepackage{graphicx}
\usepackage{bm}
\usepackage{color}
\usepackage{lineno}
\usepackage{gensymb}
\usepackage{soul}
\usepackage{textgreek}
\usepackage[utf8x]{inputenc}
\usepackage{xcolor}

\begin{document}

\title{Growth-induced phase changes in swimming bacteria at finite liquid interfaces}

\author{Blake Langeslay$^1$}
\author{Gabriel Juarez$^2$}
\thanks{Email address.}\email{gjuarez@illinois.edu}

\affiliation{$^1$Department of Physics, University of Illinois Urbana-Champaign, Urbana, Illinois 61801, USA}

\affiliation{$^2$Department of Mechanical Science and Engineering, University of Illinois Urbana-Champaign, Urbana, Illinois 61801, USA}

\date{\today}

% \linenumbers

%--------------%
%%% Abstract %%%
%--------------%
\begin{abstract}

We introduce a system of bacteria confined to a finite 2D oil-water interface and driven on two distinct time scales by motility and by growth. The combined effect of activity on different time scales creates transitions between several common collective behaviors. These transitions are observed using time-lapse microscopy with high spatial and temporal resolution over eight hours. We sequentially observe an initial dilute state, a clustered state, an active turbulent state, and a glassy state, and we are able to directly observe and characterize the transitions between these states. This system allows the investigation of emergent effects surrounding transitions between phases, expanding on studies that have considered them in isolation. In particular, a peak in the velocity correlation length is observed at the turbulent-glassy transition, suggesting that this transition significantly increases the active turbulent length scale.

\end{abstract}

\maketitle

%------------------%
%%% Introduction %%%
%------------------%
\section{Introduction}

In active matter systems, simple, small-scale interactions give rise to emergent behaviors with much higher complexity that occur at a larger scale. 
These fascinating collective behaviors motivate much of the study of active matter \cite{Gompper}. 
Emergent effects can be observed in a variety of naturally occurring systems such as concentrated bacteria suspensions \cite{Beer,Peng,Cisneros}, schooling fish \cite{Giannini},  growing epithelial cells \cite{Blanch-Mercader,Morris,Henkes2}, and even the construction of termite nests \cite{Heyde}. 
Several distinctive emergent behaviors appear across many such disparate systems. 
These include clustering \cite{Beer,Cates,Geyer,Ginot}, active turbulence \cite{Thampi1,Wensink,Dunkel,Doostmohammadi,Blanch-Mercader,Peng,Wolgemuth}, and jamming or glassy dynamics \cite{Marchetti,Angelini,Garcia}. 
The fact that these behaviors appear consistently in systems composed of wildly different elements makes their study both interesting and important to better understand a wide variety of natural processes.

Active emergent behaviors are usually studied in isolation. 
Previous work has focused on identifying which combinations of system parameters produce a single behavior in a given system \cite{Peng,Opathalage} or on how variations in system parameters affect the details of a behavior \cite{Deng,Dunkel,Thampi1}. 
Some studies have shown the presence of multiple distinct behaviors in a system, either over multiple experiments \cite{Peng,Wensink} or in different regions of the same experiment \cite{Beer}, leading to phase-space diagrams that represent the emergence of the different behaviors. 
However, these results do not focus on the transition regions between different active behaviors, where effects from multiple behaviors can be observed simultaneously. 
They also do not observe transitions between behaviors over time.

To observe such transitions, a system must be evolving on two distinctly different time scales: a shorter time scale based on the activity, and a longer time scale modifying parameters that control the type of activity. 
A natural choice of active agent is growing motile bacteria. 
Swimming-driven activity on the scale of seconds to minutes has been previously shown to drive collective behaviors such as jet and vortex formation and large-scale coherent motion \cite{Cisneros, Dombrowski, Dunkel}. 
Growth-driven activity on the scale of hours has been previously shown to produce fundamental behavioral changes when in a confined environment, including global alignment and surface deformations \cite{volfson, hickl}.

We present a system of motile \textit{E. coli} constructed to display both short-time scale collective behavior and long-time scale evolution. 
As the cells grow and their density increases over eight hours, the system transitions between dilute, clustering, turbulent, and glassy behavior in a single experiment. 
The emergent behaviors in these behaviorally distinct time intervals will be referred to as phases. 
The dual time scale of the system enables observation of transitions between phases. In the turbulent-glassy transition, velocity correlation length increases to more than triple the length in either phase individually. This indicates a unique effect of the combined phases which could not be produced by either type of collective behavior acting in isolation.

%--------------------------%
%%% Experimental methods %%%
%--------------------------%
\section{Experimental methods}

Bacteria cultures were prepared by growing \textit{E. coli} strain MG1655-motile for 12 hours in LB broth (Sigma-Aldrich, L3022-250G) in an orbital shaker at 30 $^\circ$C and 180 RPM. 
The bacteria were rod-shaped, approximately 4 \textmu m in length and 1 \textmu m in width. 
In growth media, swimming speeds of 12 \textmu m/s were observed, with 70\% of bacteria swimming.

To create and observe confined oil-water interfaces, experiments were conducted in custom cylindrical chambers on glass slides. 
These were constructed by attaching laser-cut acrylic discs to slides with epoxy resin, resulting in wells of diameter 6.6 mm and depth 6.4 mm. 
Within these wells, a thin layer of mineral oil was spread evenly on the slide. 
A copper TEM grid (SPI Supplies, 2010C-XA) with square apertures 205 microns on a side was placed on top of the oil. 
The \textit{E. coli} culture was diluted in LB broth to an OD of 0.03 -- 0.06, and 150 \textmu L of the diluted culture was added to the well. 
This resulted in the TEM grid pinning a flat interface between the oil and the culture in each of the grid's apertures. 
The well was then filled with additional LB broth and covered with a glass coverslip before being imaged.

Time-lapse phase contrast microscopy was used to image the oil-water interface over 8 hours using a $60\times$ objective. 
The resulting resolution was 0.11 \textmu m per pixel with a 2 \textmu m depth of field. 
A single aperture in the TEM grid was imaged at a time, selected for clarity and lack of visible contaminants. 
Images were taken with a scientific CMOS camera using a 50 ms exposure time. 
The image acquisition rates used were 2 frames per minute when tracking growth and 10 frames per second when tracking motility.

To quantify cell growth, the average packing fraction \(\phi(t)\) was calculated. 
This was defined as the area fraction of the interface covered by cells at a given time. 
Measurements of \(\phi\) were obtained from brightness thresholding of phase contrast images in MATLAB. 
These measurements were then normalized to match manual counts of bacteria at test frames. 
As cell growth drives the changes in collective behavior in the system, \(\phi\) is treated as the governing parameter. Other measurable quantities are reported as functions of \(\phi\) rather than time.

Spatial variation in cell arrangement was quantified using the local packing fraction \(\phi_{L}\). 
This was calculated similarly to \(\phi\) by using brightness thresholding of phase contrast images. 
Rather than taking the average value of packing fraction over the entire experimental area, the average was found for each grid square of a 40 x 40 array (resulting in 5.1 \textmu m squares).

Activity was characterized using velocity fields of cell motion. 
These were generated through particle image velocimetry (PIV), which quantified the average cell displacement between frames \cite{Thielicke}. 

%-----------------------------------------------------%
%%% Figure 1 - bacteria at oil-water interfaces %%%
%-----------------------------------------------------%
\begin{figure*}
\centering
	\includegraphics[width=\linewidth]{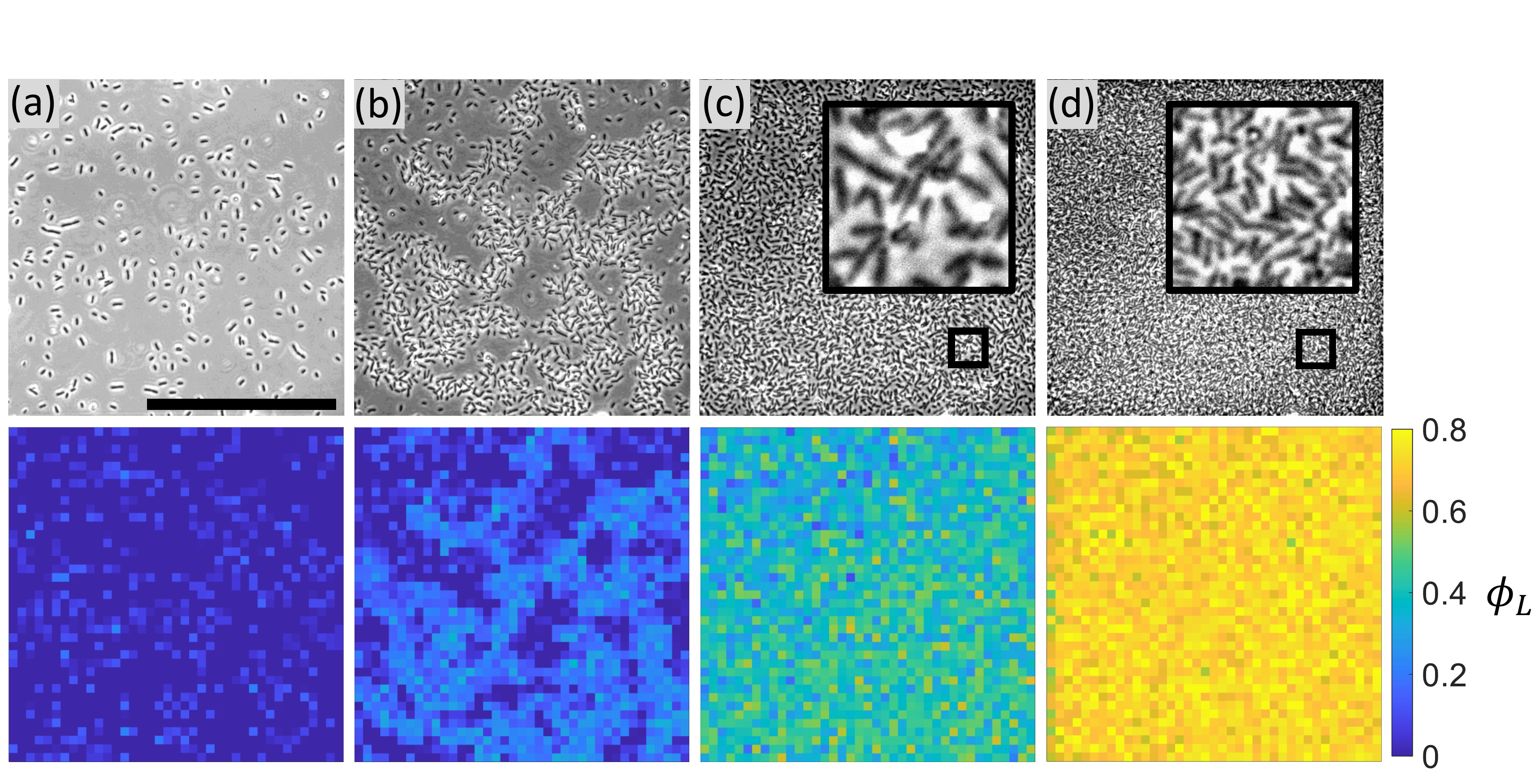}
\caption{
Bacteria growing at finite oil-water interfaces, shown in order of increasing time and packing fraction from left to right. 
(top) Phase contrast images and (bottom) local packing fraction area maps show snapshots of the different phases: (a) dilute, (b) clustered, (c) turbulent, and (d) glassy (\(\phi=0.02, 0.12, 0.28, 0.71\), respectively).
Scale bar is 100 \textmu m.
Insets are $18 \times 18$ \textmu m and show bacterial arrangement.
}
	\label{fig:figone}
\end{figure*}

%-------------%
%%% Results %%%
%-------------%
\section{Results}

Bacteria growing and swimming at the oil-water interface appear as dark rods on a light background in phase contrast images, shown in Figure \ref{fig:figone} (top row). 
Over time the confined cells grow to cover more of the available interface. 
Initially, bacteria are sparsely distributed, then separate into groups, and finally distribute homogeneously, as shown in Figure \ref{fig:figone}(a-d).
The varying spatial distributions of cells over time are apparent in heatmaps of local packing fraction \(\phi_{L}\), shown in Figure \ref{fig:figone} (bottom row).

\begin{figure*}
\centering
	\includegraphics[width=\linewidth]{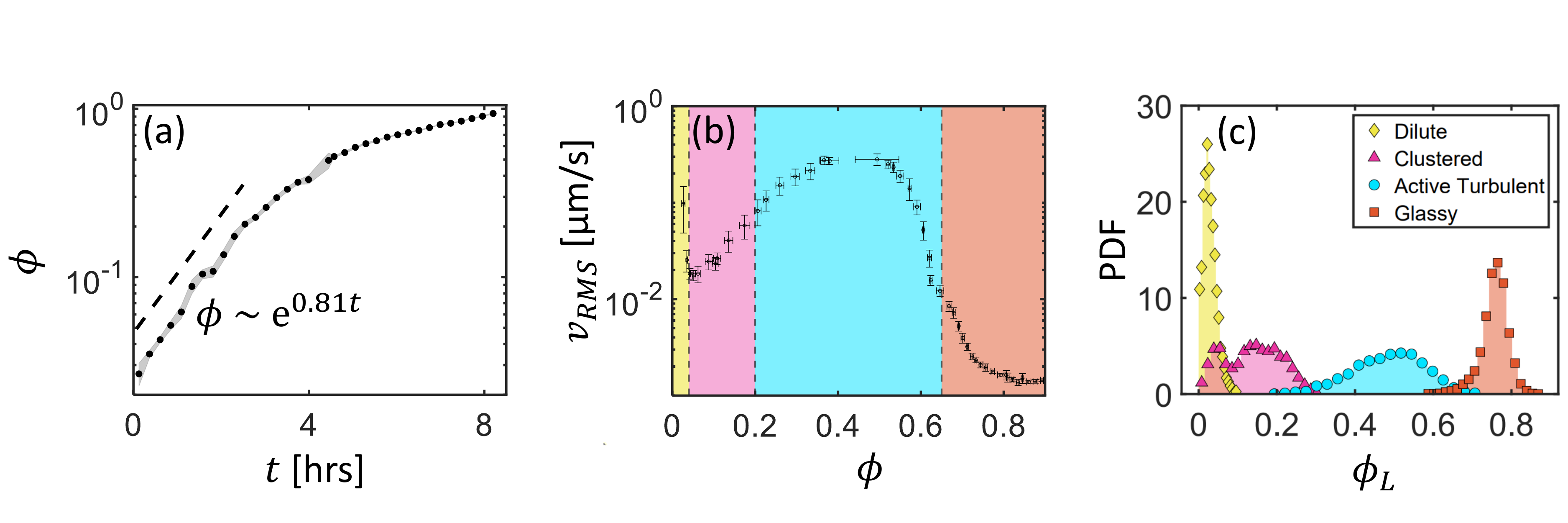}
\caption{
(a) Packing fraction evolution in growing bacteria over 8 hours. 
The dotted line is an exponential fit during the first 150 minutes. 
(b) RMS velocity of the bacterial colony as a function of packing fraction. 
Approximate boundaries between the different phases are indicated (from left to right: dilute, clustered, active turbulent, glassy). 
(c) Probability distributions of local packing fractions in each phase. 
Each distribution is taken over a 20-minute interval, with average \(\phi=0.03, 0.10, 0.30, 0.75\), respectively.
}
	\label{fig:figtwo}
\end{figure*}

The packing fraction \(\phi\) increases monotonically from less than 0.03 to approximately 0.9 as cells grow and divide at the oil-water interface over 8 hours, shown in Figure \ref{fig:figtwo}(a). 
Initially, this increase is exponential with a time constant of 0.81 hrs\(^{-1}\). 
At 150 minutes (\(\phi=0.21\)), the rate of increase of \(\phi\) decreases significantly and is no longer well described by an exponential function.

The RMS cell velocity (\(v_{RMS}=\sqrt{\langle v_{x}^{2}+v_{y}^{2} \rangle}\)) reveals that as \(\phi\) increases the activity of the system changes, as shown in Figure \ref{fig:figtwo}(b). 
The RMS velocity is non-monotonic with increasing \(\phi\). 
Instead, its evolution displays peaks and valleys spanning orders of magnitude as \(\phi\) increases, with distinct periods of high and low velocities. 
This suggests a series of transitions between phases with distinct modes of collective behavior.

%-----------------------------------------------------%
%%% Figure 2 - packing fraction, RMS velocity, & distributions %%%
%-----------------------------------------------------%

A series of behavioral phases were identified over the system's evolution. 
In chronological order, these are: dilute, clustered, turbulent, and glassy. 
Visual differences between phases can be observed in the phase contrast images and local packing fraction heatmaps, shown in Figure \ref{fig:figone}. 
Dilute bacteria are arranged with no ordering and little to no contact, clustered bacteria segregate into high- and low-density regions, and turbulent and glassy bacteria distribute homogeneously with no visible structure. 
The distinction between the turbulent and glassy phases can be visually identified where turbulent cells are highly active (shown in supplementary videos SV3 and SV4) and glassy cells are predominantly stationary (supplementary video SV5).
The approximate boundaries of the phases are shown in the colored regions of Figure \ref{fig:figtwo}(b).

%--------%
% DILUTE %
%--------%

%% FIGURE 1a %%

The initial dilute state at the lowest cell densities ($\phi<0.04$) exhibits no collective behavior. 
Adsorbed bacteria are uniformly spaced and randomly oriented as shown by phase contrast images and local packing fractions in Figure \ref{fig:figone}(a). 
In this regime, near-circular ``curly'' swimming trajectories are observed in some bacteria, as previously seen for motile bacteria at liquid interfaces \cite{Morse, Deng}. See supplementary Figure S1.

%-----------%
% CLUSTERED %
%-----------%

%% FIGURE 1b %%

Next, as the packing fraction increases with cell growth ($0.04<\phi<0.2$), a clustered phase emerges. 
Here the formation of large clusters of cells is seen in phase contrast images and heatmaps of $\phi_{L}$ in Figure \ref{fig:figone}(b). 
Bacteria accumulate in high-density regions with populations ranging between ten and several hundred, leaving a low-density background of isolated bacteria. 
Concurrently with the formation of clusters, the RMS cell velocity drops by nearly an order of magnitude, shown in Figure \ref{fig:figtwo}(b).

%% FIGURE 2c %%

The difference between high- and low-density regions produced by clustering is quantified in the distributions of local packing fractions, shown in Figure \ref{fig:figtwo}(c). 
These display a bimodal distribution of $\phi_{L}$ in the clustered phase, similar to previously observed clustering behavior in swarming bacteria \cite{Beer}. 
All other phases have single-peaked distributions of $\phi$, indicating that cells in those phases are homogeneously distributed. 
The two peaks of the $\phi_{L}$ distribution in the clustered phase represent a high local packing fraction within the clusters and a low background packing fraction. 
The lower-density peak of the clustered phase coincides with the single peak of the dilute phase, suggesting that non-clustered background bacteria continue to behave as if in the dilute state.

%------------------%
% ACTIVE TURBULENT %
%------------------%

%-----------------------------------------------------%
%%% Figure 3 -  %%%
%-----------------------------------------------------%
\begin{figure*}
\centering
	\includegraphics[width=\linewidth]{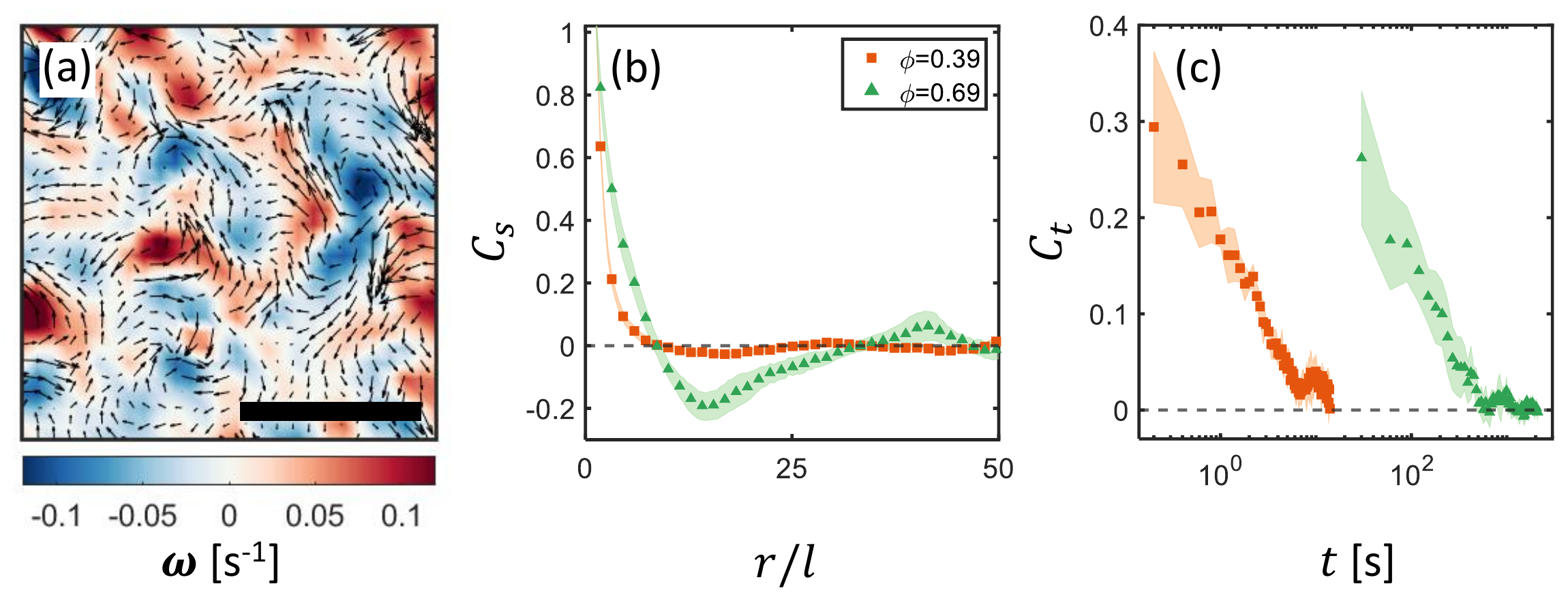}
\caption{
(a) Velocity field with a vorticity heatmap in the turbulent phase with $\phi=0.39$. 
Scalebar is 40 \textmu m.
(b) Spatial velocity correlation functions of cells in the turbulent phase ($\phi=0.39$, orange squares) and during the turbulent-glassy transition ($\phi=0.69$, green triangles). 
The distance $r$ is normalized by the bacteria length $l$. 
(c) Temporal velocity correlations of cells in the turbulent phase ($\phi=0.39$, orange squares) and during the turbulent-glassy transition ($\phi=0.69$, green triangles).
}
	\label{fig:figthree}
\end{figure*}

%%FIGURE 1c and 3a %%

At $\phi=0.2$, the system enters the active turbulent phase. 
Here clusters break up, leading to a homogeneous distribution of positions shown in Figure \ref{fig:figone}(c). 
The packing fraction distribution returns to a single-peaked form, shown in Figure \ref{fig:figtwo}(c). 
Cells move in locally coherent flows, circulating in shifting patterns of vortices throughout reminiscent of classical inertial turbulence shown in Figure \ref{fig:figthree}(a). 
The RMS velocity of these flows is much higher than that of the cells in the clustered phase, returning to a range similar to that of the dilute state shown in Figure \ref{fig:figtwo}(b).

To characterize the flow in this phase, the spatial and temporal correlation functions of velocity \(C_{s}(t,r)\) and \(C_{t}(t,r)\), common measures of active turbulence, were calculated \cite{Thampi1,Wensink,Dunkel}. 
The spatial correlation measures the coherence of the flow at varying distances, while the temporal correlation measures how quickly the flow structure changes in time. 
The spatial velocity correlation function at a time \(t\) is calculated as follows:
\begin{equation} 
C_{s}(t,r)= \frac{ \langle \vec{v_i}(t) \cdot \vec{v_j}(t) \rangle}{\langle v^2 \rangle}
\end{equation} 
where the expectation value is calculated over all pairs of points \(i\) and \(j\) separated by a distance \(r\). The temporal velocity correlation function is similarly:
\begin{equation} 
C_{t}(t,\delta t)= \frac{ \langle \vec{v_i}(t) \cdot \vec{v_i}(t+\delta t) \rangle}{\langle v^2 \rangle}
\end{equation} 
where the expectation value is calculated over each point \(i\).

%% FIGURE 3b

Representative spatial correlation functions in the turbulent phase and near the turbulent-glassy transition are shown in Figure \ref{fig:figthree}(b). 
In both cases, the correlation function drops to zero after a finite number of bacteria lengths ($\approx 10$), much less than the total system size. 
Past this point, both functions become negative, showing anticorrelated velocity at higher distances. 
This reflects the fact that points on opposite sides of a vortex move in opposite directions. 
The correlation has higher magnitudes - both positive and negative - during the turbulent-glassy transition ($\phi=0.69$), showing more consistently coherent flow at the higher packing fraction.

%% FIGURE 3c %%

Temporal correlation functions for the same two cases are shown in Figure \ref{fig:figthree}(c). 
Both functions fall to zero after a finite time that is much less than the cell growth time scale of the system.
This shows that the flow structure is constantly changing rather than being a static array of vortices, as normal for turbulence.
However, the temporal correlation function in the turbulent-glassy transition falls to zero much faster than the function in the main turbulent phase (hundreds compared to tens of seconds), reflecting a much more consistent flow pattern.

%%FIGURE 4a %%

To quantify the length scale of the system’s active turbulence, a correlation length $\xi$ was obtained from an exponential fit of the spatial correlation function \cite{Rabani}:
\begin{equation}
C_{s}(t,r)\approx \exp(-r/\xi(t))
\end{equation}
The exponential fit was performed over the domain of the correlation function up to its first zero. 
The correlation length $\xi(t)$ represents the approximate length scale over which bacteria velocities remain correlated at a given time. 

\begin{figure*}
\centering
	\includegraphics[width=\linewidth]{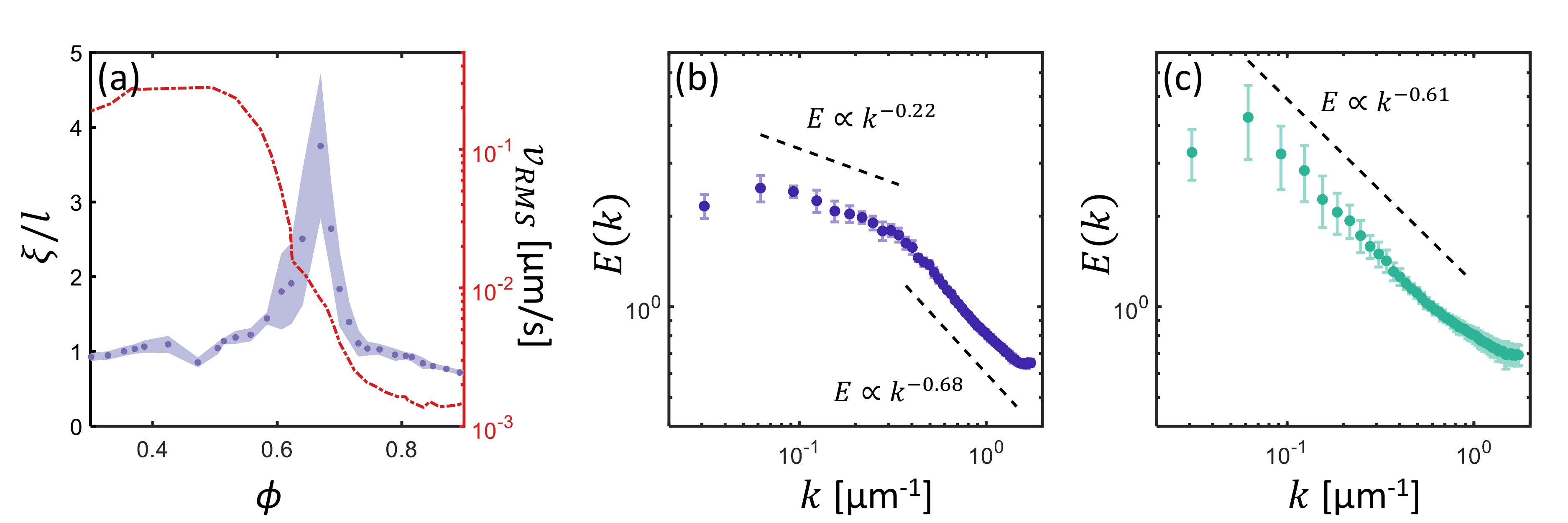}
\caption{
(a) Spatial velocity correlation length \(\xi\) normalized by bacteria length \(l\), as a function of packing fraction \(\phi\). 
Colored region represents standard error. Overlaid dashed line shows RMS velocity.
(b) Energy spectrum E(K) for \(\phi=0.51\) in the active turbulent regime. 
(c) Energy spectrum for \(\phi=0.69\) in the transition from active turbulence to glassy dynamics. 
Error bars are standard error for both energy spectra.
}
	\label{fig:figfour}
\end{figure*}

The evolution of $\xi$ with increasing packing fraction during the transition from turbulent to glassy behavior is shown in Figure \ref{fig:figfour}(a).
During the turbulent phase, $\xi \approx l$, indicating correlated motion on the length scale of a single cell. During the transition between phases, from approximately $\phi=0.65$ to $\phi=0.7$, $\xi$ temporarily increases to $\sim 4l$ before dropping back to $\sim l$ in the glassy phase.

%-----------------------------------------------------%
%%% Figure 4 -  %%%
%-----------------------------------------------------%

%% FIGURE 4b %%

To quantify the relative strength of turbulent flow at differing length scales, the energy spectra \(E(K)\) were calculated from the Fourier transforms of the spatial velocity correlation functions, shown for two different packing fractions in Figure \ref{fig:figfour}(b, c). 
At packing fractions within the active turbulent phase ($\phi=0.51$), two scaling regimes were observed: a regime with scaling exponent \(\alpha=-0.68\) at small length scales, and a regime with a lower-magnitude exponent \(\alpha=-0.22\) at large length scales. 
These correspond to the energy scaling below and above the typical vortex length scale, respectively. 
At the transition to glassy dynamics ($\phi=0.69$), the spectrum instead follows a single scaling law with an exponent \(\alpha=-0.61\).

%--------%
% GLASSY %
%--------%

%% FIGURE 1d %%

At \(\phi=0.65\) the bacteria's motion arrests at the onset of the glassy phase. 
This transition is characterized by a decrease of over an order of magnitude in the RMS velocity, toward a limiting velocity near \(10^{-3}\) \textmu m/s, as shown in Figures \ref{fig:figtwo}(b) and \ref{fig:figfour}(a). 
This is similar to the behavior observed at high densities in other bacterial systems \cite{Rabani,Beer}. 
Phase contrast images show that the bacteria are in near-constant contact with each other, as seen in Figure \ref{fig:figone}(d, inset). 

%----------------%
%%% Discussion %%%
%----------------%
\section{Discussion}

In summary, growing bacteria at a finite oil-water interface form an active system that progresses through four distinct phases (dilute, clustered, turbulent, and glassy) over the course of a single experiment due to the combination of motility-generated activity and growth-generated time evolution. 
These phases match other modes of behavior observed elsewhere in active matter. 
Because our system allows observation of the transitions between these phases, it enables novel investigation of how these behaviors emerge, interact, and dissipate.

The system's phase changes are produced by its combination of motility- and growth-based time scales. 
Motility-powered active behaviors occur on the order of seconds. 
Meanwhile, growth increases the global packing fraction on the order of hours. 
These packing fraction changes slowly modify the system's motility-powered short-time scale emergent behavior, leading to changing behavioral phase. 
The final transition to a glassy phase is enabled by the system's confinement. 
Because the cells grow on a 2D interface bounded on all four sides, the area available for growth is finite. 
This causes cells at packing fractions above \(\phi=0.65\) to interfere with each other's motility through volume exclusion and produce a glassy phase.

As \(\phi\) is directly proportional to the cell count, its doubling time is expected to match the doubling time of the cells. 
The doubling time of \(\phi\) is 51 minutes, while the strain's previously described doubling time on rich media is 33.8 minutes \cite{Fraebel}. 
This disparity likely exists because the oil-water interface provides a less optimal environment for growth, as part of an adsorbed cell's surface contacts oil rather than growth media and therefore does not contribute to nutrient intake.

The clustered phase of our system matches clustering behavior observed elsewhere in bacteria swarming on agar, where similar bimodal density distributions have been reported \cite{Beer}. 
The accumulation of bacteria in high-density regions is of particular interest because of its impact on biofilm formation \cite{Grobas}. 
In our system, increasing packing fraction in the range of $0.03-0.2$ leads to phase separation into high-density clusters and a dilute background. 
However, further increases in packing fraction cause the phase separation to break down. 
The breakdown of clustering in our system is caused not by jamming (as observed for swarming cells on agar) but by a drastic increase in particle velocity \cite{Beer}. 
This happens as higher local packing fractions lead to active collective motion, likely due to the combined hydrodynamics of dense swimming bacteria causing the system to transition to the active turbulent phase.\cite{Dombrowski,Aranson}.

Active turbulence is a widely studied phenomenon in active matter, bacterial and elsewhere \cite{Thampi1,Wensink,Dunkel,Doostmohammadi,Blanch-Mercader,Peng,Wolgemuth,Aranson}. 
The ``turbulent'' phase observed here is not a perfect match for this phenomenon, as it only weakly displays the characteristic sub-zero drop in \(C_{s}\), as shown in Figure \ref{fig:figthree}(b) \cite{Thampi1,Wensink}. 
However, it has other clear similarities, particularly its structure of vortices, as well as the fact that its correlation functions of both time and space are nonzero at short ranges but approach zero on a scale much less than that of the system, as shown in Figure \ref{fig:figthree}(b and c). 
This indicates a chaotic structure in time and space strongly reminiscent of classical inertial turbulence. 
Given this similarity and the fact that the behavior occurs in an active system, we believe that active turbulence is a good description of this phase's behavior.

The turbulent-glassy transition significantly intensifies the correlation of active turbulent motion, as per the peak in \(\xi\) at the phase transition between $0.65 < \phi < 0.7$ shown in Figure \ref{fig:figfour}.
This is remarkably similar to effects observed in swimming spherical bacteria at an air-water interface \cite{Rabani} and in growing epithelial cells \cite{Garcia}. 
Our data builds on these observations, showing that the increase in correlation length is closely confined to the transition region, with \(\xi\) nearly constant both before and after the transition. 
This implies that the increase in correlation length here is not an effect solely due to a glass transition, but an emergent phenomenon resulting from the combination of active turbulent and glassy behaviors.

In the active turbulent regime just before this transition, the energy spectrum displays two different scalings at high and low length scales. 
This matches previous work on active turbulence, where it has been found that energy scales with a higher-magnitude exponent at length scales below the typical vortex size, where energy is injected \cite{Alert,MartinezPrat}. 
Our exponents (\(\alpha=-0.68\) below the typical vortex size, \(\alpha=0.22\) above) do not match the previously reported exponents for active turbulent systems (\(\alpha=-4\) and \(-1\), respectively). 
A major difference between the model that produced these exponents and our system is that the model is of an active nematic, while bacteria in our system do not exhibit strong nematic ordering. 
This is likely related to the difference in scaling behavior.

During the transition to glassy dynamics, the observed energy spectrum changes to a form with a single scaling exponent (\(\alpha=-0.61\), close to the sub-vortex-size exponent before the transition. 
This likely occurs because the drastic increase in correlation length produces a typical vortex size that is greater than the length scale of confinement, such that the entire experiment takes place in the ``smaller than vortex'' regime. 
As such, we expect that experiments with much a larger confinement area would produce an energy spectrum with the previous dual-scaling form at these packing fractions, with the cutoff between the two scalings (the vortex size) at a much higher length scale.

The vast majority of bacteria in the glassy phase become extremely slow (~\(10^{-3}\) \textmu m/s, shown in  Figure \ref{fig:figtwo}(b). 
However, isolated higher-velocity bacteria still exist as ``caged'' particles in open pockets of the larger glassy structure \cite{Segall}. 
These caged particles have uncorrelated velocities, as they are separated by what are essentially solid walls. 
These uncorrelated velocities are much higher than the very slow correlated motion of the active glass, leading to an average low velocity correlation length for the glassy regime as shown in Figure \ref{fig:figfour}(a).

Our system's unique characteristics stemming from its combination of motility- and growth-based activity in a wet environment enable direct observation of transitions between active emergent behaviors. 
This has allowed the characterization of several interesting and novel phenomena in areas of interest in active matter such as active turbulence and clustering. 
Further research on bacteria at similar interfaces is promising, particularly utilizing differing motility and bacterial geometry. 
Additionally, numerical modeling of the active turbulent-glassy transition would be of great interest to capture the increase in length scale shown by the correlation length. 
We believe that growing bacteria on an oil-water interface are a useful model system for the investigation and direct observation of phase transitions in active matter.

%---------------------%
%%% Acknowledgments %%%
%---------------------%
\section{Acknowledgements}
 
We thank Seppe Kuehn and Karna Gowda for providing the \textit{E coli.} strain and for helpful discussions regarding cell culturing techniques.

%------------------%
%%% bibliography %%%

\end{document}